\documentclass[aps,twocolumn,showpacs]{revtex4}
\usepackage{graphicx}
\usepackage{amssymb}
\usepackage{amsmath}
\newcommand{\be}{\begin{equation}}
\newcommand{\ee}{\end{equation}}
\newcommand{\ben}{\begin{eqnarray}}
\newcommand{\een}{\end{eqnarray}}
\newcommand{\bes}{\begin{subequations}}
\newcommand{\ees}{\end{subequations}}
\newcommand{\bb}{\bibitem}

\newcommand{\bfi}{\begin{figure}}
\newcommand{\efi}{\end{figure}}
\newcommand{\bc}{\begin{center}}
\newcommand{\ec}{\end{center}}

\begin{document}
\title{On the presence of twinlike models in cosmology}
\author{D. Bazeia$^{1,2}$ and J.D. Dantas$^{1,3}$}
\affiliation{$^1$Departamento de F\'{\i}sica, Universidade Federal da Para\'{\i}ba, 58051-970 Jo\~ao Pessoa, PB, Brazil}
\affiliation{$^2$Departamento de F\'{\i}sica, Universidade Federal de Campina Grande, Campina Grande, PB, Brazil}
\affiliation{$^3$Unidade Acad\^emica de Educa\c c\~ao, Centro de Educa\c c\~ao e Sa\'ude, Universidade Federal de Campina Grande, 58175-000 Cuit\'e, PB, Brazil}
\date{\today}

\begin{abstract}
We study cosmological models described by a single real scalar field.
We work within the first-order framework, and we show how the first-order equations simplify the investigation,
leading to a direct search of twinlike theories. The procedure is used to introduce distinct models that support the same
first-order equations, with the very same energy densities and pressure in flat spacetime. The presence of curvature forbids
the construction of twinlike models in the cosmological scenario here investigated.
\end{abstract}

\pacs{98.80.Cq}

\maketitle


\section{Introduction}

Defect structures appear in distinct areas of nonlinear science. In high energy physics, the most known defects are kinks, vortices and monopoles. Kinks are the simplest structures, and they appear in one spatial dimension, in the presence of a single real scalar field \cite{book1,book2}.  In the recent works \cite{altw,jp1,sc1,jp2,sc2,jp3}, interesting novelties on defect structures have been stablished, concerning the presence of twinlike models, which are models describing distinct field theories but sharing the very same defect structures. 

In the present work we go further on the subject, investigating relativistic systems described by real scalar field in Friedmann-Robertson-Walker (FRW) cosmology. The subject has been investigated in a diversity of contexts, but here we shall bring novelties related to the issue recently studied in \cite{altw,jp1,sc1,jp2,sc2,jp3}, concerning the presence of twinlike models. We shall take advantage 
of the formalism introduced in Ref.~\cite{first}, from which it is possible to get to first-order differential equations that solve the equations of motion for a general scalar field model. The procedure is valid in flat and in curved spacetime. We use this in the case of standard FRW cosmology, and also for the scalar field being tachyonic.  In the process, we further ask for models that present the very same first-order equations, with the same energy density and pressure, and we name them twinlike models. 

We study the subject in flat and in curved spacetime, and we organize the work as follows: In Sec.~\ref{sec2} we explore the standard and tachyonic models, and we comment on some specific properties the models engender, in particular on how to introduce twinlike models. In Sec.~\ref{sec4} we illustrate the results investigating a specific example. In Sec.~\ref{sec5} we deal with curved geometry, and we end the work in Sec.~\ref{sec6}, summarizing the results.


\section{Twinlike models}
\label{sec2}

We start with the Einstein-Hilbert action
\be
S=\int d^4x\sqrt{-g}\left[-\frac{1}{4}R+{\cal L}(\phi,\partial_\mu\phi)\right],
\ee
where $4\pi G=1$, $g$ is the determinant of the metric, $R$ is the scalar curvature and ${\cal L}$ is the Lagrange density that describes the scalar field model.

In this general scenario, Einstein's equations are given by
\be
G_{\mu\nu}=2T_{\mu\nu},
\ee
where
\be
G_{\mu\nu}=R_{\mu\nu}-\frac{1}{2}g_{\mu\nu}R
\ee
is the Einstein tensor and
\be
T_{\mu\nu}=2\frac{\partial{\cal L}}{\partial g^{\mu\nu}}-g_{\mu\nu}{\cal L}
\ee
is the energy-momentum tensor.

The metric we use is the standard FRW metric, which has the form
\be
ds^2=dt^2-a^2(t)\left(\frac{dr^2}{1-kr^2}+r^2d\Omega^2\right).
\ee
where $a$ is the scale factor, $r$ is the radial coordinate and $d\Omega^2$ describes the angular portion of the metric. Also, $k=0$ for flat geometry, and $k=\pm1$ for spherical (+) or hyperbolic (--) geometry.
The Friedmann equations have the form
\be
H^2=\frac{2}{3}\rho-\frac{k}{a^2}
\ee
and
\be
\frac{\ddot{a}}{a}=-\frac{1}{3}(\rho+3p),
\ee
where $H={\dot a}/a$ is the Hubble parameter, and $\rho$ and $p$ are energy  density and pressure. We obtain
\be
\dot{\rho}+3H(\rho +p)=0,
\ee
and the acceleration parameter has the form
\be
q=\frac{\ddot{a}a}{\dot{a}^2}=1+\frac{\dot{H}}{H^2}.
\ee

\subsection{Standard case}

Let us now consider the scalar field described by standard kinematics. We have
\be
{\cal L}=\frac{1}{2}\partial_{\mu}\phi\,\partial^{\mu}\phi -V(\phi).
\ee
where $V(\phi)$ is the potential of the scalar field. In this case  energy density and pressure are given by
\bes\ben
\rho&=&\frac{1}{2}\dot{\phi}^2+V(\phi),
\\
p&=&\frac{1}{2}\dot{\phi}^2-V(\phi),
\een\ees
and the scalar field obeys
\be
\ddot{\phi}+3H\dot{\phi}+V_{\phi}=0.
\ee
In the case of flat geometry, for $k=0$ we have
\be
H^2=\frac{1}{3}\dot{\phi}^2+\frac{2}{3}V.
\ee

\subsection{Tachyonic case}

If the scalar field is tachyonic, we change kinematics as follows
\be
{\cal L}=-U(\phi)\sqrt{1-\partial_{\mu}\phi\,\partial^{\mu}\phi}+f(\phi),
\ee
where $U(\phi)$ and $f(\phi)$ are functions to be determined. In this case,
\bes\ben
\rho&=&\frac{U}{\sqrt{1-\dot{\phi}^2}}-f,
\\
p&=&-U\sqrt{1-\dot{\phi}^2}+f.
\een\ees
and the scalar field obeys
\be
\ddot{\phi}+\left(1-\dot{\phi}^2\right)\left(3H\dot{\phi}+\frac{U_{\phi}}{U}\right)-\left(1-\dot{\phi}^2\right)^{3/2}\frac{f_{\phi}}{U}=0.
\ee

In the case of flat geometry, for $k=0$ we get
\be
H^2=\frac{2U}{3\sqrt{1-\dot{\phi}^2}}-\frac{2}{3}f.
\ee

\subsection{First-order formalism}
\label{sec3}

Let us now consider the case of flat geometry. Here we follow Ref.~\cite{first}, and for $k=0$ we take $H=W(\phi)$, so we get
\be
W_{\phi}\dot{\phi}=-(\rho+p).
\ee

In standard case, we have
\be
\dot{\phi}=-W_{\phi}
\label{um}
\ee
and
\be
V=\frac{3}{2}W^2-\frac{1}{2}W_{\phi}^2.
\ee

In the case of a tachyonic field, we also use $H=W(\phi)$, but now we have
\be
\dot{\phi}=-\frac{W_{\phi}}{\frac{3}{2}W^2+f}
\label{dois}
\ee
and
\be
U=\sqrt{1-\dot{\phi}^2}\left(\frac{3}{2}W^2+f\right).
\ee

In order to get to twinlike models, we make \eqref{um} equal to \eqref{dois}. Here we get
\be
f=1-\frac{3}{2}W^2.
\ee
Thus,
\be
U=\sqrt{1-W_{\phi}^2},
\ee
and the Lagrange density of the twin tachyonic model has the following form
\be
{\cal L}=-\sqrt{1-W_{\phi}^2}\,\sqrt{1-\partial_{\mu}\phi\partial^{\mu}\phi}+1-\frac{3}{2}W^2.
\ee

We note that in both standard and tachyonic cases, we have the same energy density
\be
\rho=\frac{3}{2}W^2,
\ee
and the same pressure
\be
p=W_{\phi}^2-\frac{3}{2}W^2.
\ee

Also, the acceleration parameter takes the same form
\be
q=1-\frac{W_{\phi}^2}{W^2}.
\ee
for both cases, with standard and tachyonic kinematics.

\section{Example}
\label{sec4}

As an example we take $W=A\cosh(B\phi)$. So,
\be
V=\frac{3}{2}A^2\cosh^2(B\phi)-\frac{1}{2}A^2B^2\sinh^2(B\phi)
\label{vpotum}
\ee
and
\be
U=\sqrt{1-A^2B^2\sinh^2(B\phi)}.
\label{upotum}
\ee
The two cases are depicted in the Fig. \ref{stpotum}.

\bfi[h!]
\bc
\includegraphics[scale=1.3]{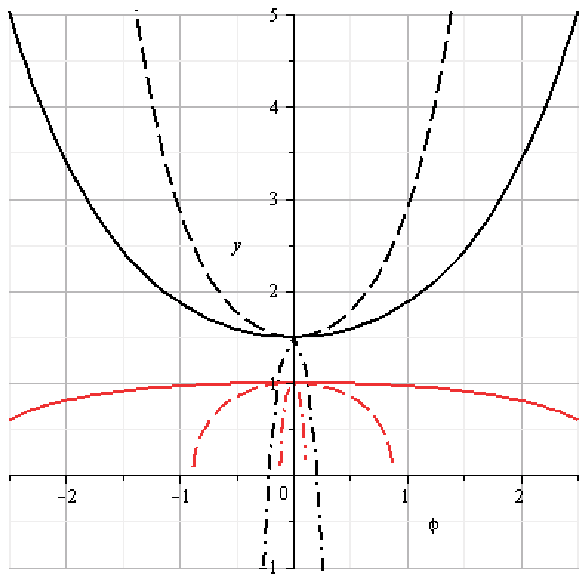}
\caption{Plots of the potentials (\ref{vpotum}) (black, upper curves) and (\ref{upotum}) (red, lower curves) for $A=1$ and $B=0.5,1$ and $3$, corresponding to solid, dashed and dot-dashed curves, respectively.}
\label{stpotum}
\ec
\efi

The solution of the equation of motion is
\be
\phi=\frac{1}{B}\ln\left(\tanh\frac{AB^2t}{2}\right).
\ee
Here the Hubble parameters is given by
\be
H=A\cosh\left[\ln\left(\tanh\frac{AB^2t}{2}\right)\right]\, .
\label{hum}
\ee
Also, the acceleration parameter $q$ has the form
\be
q=1-\frac{B^2{\rm sech}^4\left(\frac{AB^2t}{2}\right)}{\left[\tanh^2\left(\frac{AB^2t}{2}\right)+1\right]^2},
\ee
which is depicted In Fig.~\ref{stqum}, where the solid curve indicates an always accelerated expansion. The dashed curve shows an expansion that begins to accelerate at $t=0$,
and the dot-dashed curve indicates evolution from decelerated to accelerated phase.

\bfi[h!]
\bc
\includegraphics[scale=1.3]{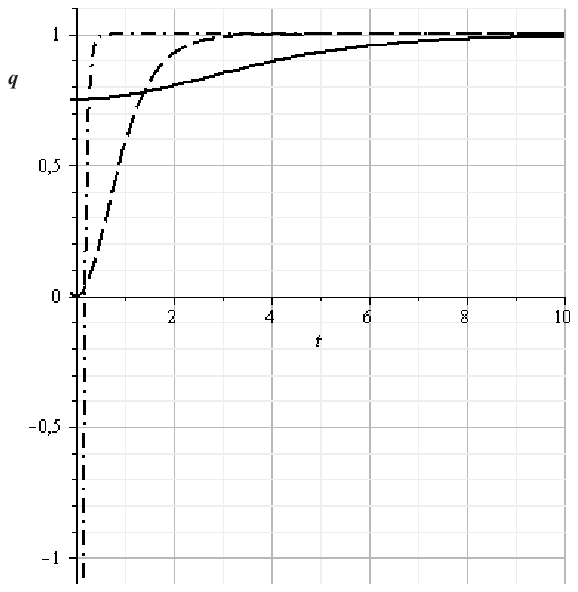}
\caption{Acceleration parameter $q$ for $A=1$ and for $B=0.5,1$ and $3$, corresponding to solid, dashed and dot-dashed curves, respectively.}
\label{stqum}
\ec
\efi

\section{Presence of Curvature}
\label{sec5}

In the presence of curvature, we consider the case of a non vanishing $k$. If we insist with $H=W$, as in Ref.~\cite{first}, we have
\be
W_{\phi}\dot{\phi}=-(\rho+p)+\frac{k}{a^2}.
\ee
We introduce $\alpha$ as a real number and $Z=Z(\phi)$ as a function of the scalar field. In this case, the choice ${1}/{a^2}=\alpha Z\dot{\phi}$ leads us to, in the standard case,
\be
\dot{\phi}=k\alpha Z-W_{\phi}.
\label{ppu}
\ee
The potential is given by
\be
V=\frac{3}{2}W^2+(k\alpha Z-W_{\phi})\left(k\alpha Z-\frac{1}{2}W_{\phi}\right).
\ee
Here the new function $Z$ is constrained to obey
\be\label{cs}
W_{\phi\phi}Z+W_{\phi}Z_{\phi}-2k\alpha ZZ_{\phi}-2WZ=0.
\ee

In the case of the tachyonic field, we obtain
\be
\dot{\phi}=-\frac{W_{\phi}}{\frac{3}{2}\left(W^2+\frac{k}{a^2}\right)+f}.
\label{ppd}
\ee
Taking (\ref{ppu}) equal to (\ref{ppd}), we determine $f$; it is given by
\be
f=-\frac{3}{2}W^2-\frac{3}{2}k\alpha Z(k\alpha Z-W_{\phi})-\frac{W_{\phi}}{k\alpha Z-W_{\phi}}.
\ee
The potential is then
\be
U=-\frac{W_{\phi}\sqrt{1-(k\alpha Z-W_{\phi})^2}}{k\alpha Z-W_{\phi}}.
\ee

Here we note that the above expressions for ${\dot\phi}$, for the potentials $V(\phi)$ and $U(\phi)$ and for $f(\phi)$, they all reduce to the corresponding equations of the flat case when one sets $k=0$. 
In the tachyonic case, however, the constraint that appears for $k\neq0$ due the the presence of the new function $Z(\phi)$ is given by
\be
W_{\phi\phi}Z+W_{\phi}Z_{\phi}-2k\alpha ZZ_{\phi}=0.
\ee
The above constraint is different from the constraint of the standard case, as given by Eq.~\eqref{cs}. This result suggests
that the presence of curvature $(k\neq0)$ distinguishes the two models, forbidding the construction of twinlike models for
the standard and tachyonic cases.

\section{Final comments}
\label{sec6}

In this work we studied the presence of twinlike models in FRW cosmology driven by a single real scalar field, in flat and in curved spacetime.
We showed that it is possible to have models driven by standard and tachyonic dynamics, sharing the same
first-order equations, the same energy density and the same pressure, but only in flat spacetime. The presence of curvature
forbids the construction of twinlike models for the scalar field being driven by standard and tachyonic dynamics,
within the FRW cosmological scenario.

\acknowledgments

We would like to thank CAPES and CNPq for partial financial support.

\end{document}